\documentclass[aps,prl,twocolumn,superscriptaddress,showpacs]{revtex4}
\usepackage{lipsum}
\usepackage{amsmath}
\usepackage[utf8]{inputenc}
\usepackage{pdfpages}
\usepackage{graphicx}
\usepackage{physics}
\usepackage{amsmath}
\usepackage{epstopdf}
\usepackage{amssymb}
\usepackage{mathtools}
\usepackage[colorlinks]{hyperref}
\AtBeginDocument{%
  \hypersetup{
    citecolor=blue,
    linkcolor=blue,
    urlcolor=blue}}

\begin{document}

\title{Efficient quantum communications with multiplexed coherent state fingerprints}

\author{Niraj Kumar} \affiliation{LTCI, CNRS, T\'el\'ecom ParisTech, Universit\'e Paris-Saclay, 75013 Paris, France}\affiliation{IRIF, CNRS, Universit\'e Paris Diderot, Sorbonne Paris Cit\'e, 75013 Paris, France}
\author{Eleni Diamanti}  \affiliation{LTCI, CNRS, T\'el\'ecom ParisTech, Universit\'e Paris-Saclay, 75013 Paris, France}\affiliation{LIP6, CNRS, Universit\'e Pierre et Marie Curie, Sorbonne Universit\'es, 75005 Paris, France}
\author{Iordanis Kerenidis}  \affiliation{IRIF, CNRS, Universit\'e Paris Diderot, Sorbonne Paris Cit\'e, 75013 Paris, France}\affiliation{CQT, National University of Singapore, Singapore}
\date{\today}

\begin{abstract}
We provide the first example of a communication model and a distributed task, for which there exists a realistic quantum protocol which is asymptotically more efficient than any classical protocol, both in the communication and the information resources. For this, we extend a recently proposed coherent state mapping for quantum communication protocols, introduce the notion of multiplexed coherent state fingerprints and show how to use them to design an efficient quantum protocol for estimating the Euclidean distance of two real vectors within a constant factor.
\end{abstract}

\maketitle

Quantum information processing harnesses the power of quantum mechanics in order to enhance the efficiency and security of information and communication technologies.
The main goal is to find tasks for which it is possible to prove theoretically the superiority of quantum information compared to classical information and to verify it experimentally. This is illustrated, for instance, in nonlocal games, where experiments have confirmed the violation of Bell inequalities that correspond to the CHSH and other games \cite{hensen2015loophole,matsukevich2008bell,ansmann2009violation,pappa2015nonlocality}. Another prominent example is quantum cryptography, where many protocols with unconditionally stronger security than classically possible have been demonstrated, including quantum key distribution, digital signatures or coin flipping \cite{scarani2009security,donaldson2016experimental,pappa2014experimental}. Unlike the above mentioned cases, most quantum algorithms are far from been implementable with current technologies, with the exception of non-universal boson sampling machines that have been realized for small inputs \cite{tillmann2013experimental,crespi2013integrated,spagnolo2014experimental}.

Communication complexity is an ideal model for testing quantum mechanics and for understanding the efficiency of quantum networks. This model studies the amount of communication required by separate parties to jointly compute a task. There are several examples where communicating quantum information can result in considerable savings in the communication overhead \cite{buhrman2001quantum,buhrman1998quantum,raz1999exponential,bar2004exponential,gavinsky2007exponential,gavinsky2016entangled,regev2011quantum}.
Nevertheless, it is in general difficult to test these results experimentally and demonstrate quantum superiority in practice since the quantum protocols typically necessitate large, highly entangled states, which are out of reach of current photonic technologies.

Recently, Arrazola and L\"utkenhaus proposed a mapping for encoding quantum communication protocols involving pure states of many qubits, unitary operations and projective measurements to protocols based on coherent states of light in a superposition of optical modes, linear optics operations and single-photon detection \cite{arrazola2014coherent}. This powerful model was used to propose the practical implementation of coherent state quantum fingerprints \cite{arrazola2014quantum}, leading to two experimental demonstrations: a proof-of-principle use of such fingerprints for solving the communication task of Equality asymptotically better than the best known classical protocol with respect to the transmitted information \cite{xu2015experimental}; and a subsequent implementation beating the classical lower bound for the transmitted information \cite{guan2016observation}. Following these demonstrations that have focused on Equality and on transmitted information, an important question remains: is there a realistic model for proving and testing in practice that quantum information is asymptotically better than classical for communication tasks with respect to \emph{all} important communication and information resources?

We answer in the affirmative by proposing a communication model and a distributed task for which we prove that quantum mechanics allows for a considerably more efficient protocol in all relevant resources. We do this by building upon the mapping of \cite{arrazola2014coherent} to introduce multiplexed coherent state fingerprints and show how to use them for solving efficiently a task that is at the foundation of many applications in Machine Learning, namely estimating the Euclidean distance of two real vectors within a constant factor. Our results show that, in principle, it is possible to demonstrate quantum superiority for advanced communication tasks in quantum networks using photonic technologies within experimental reach.

\emph{Communication resources}. We start by defining the simultaneous message passing model and the resources that we are trying to optimize. In this model, two players, Alice and Bob, receive inputs $x$ and $y$ respectively from an input set $\mathcal{X}\times\mathcal{Y}$. Their task is to use some private coins and send a single and smallest possible message to a Referee, who should be able to compute the value of a function $f(x,y)$ with a small error $\delta \in \{0,1\}$.

The communication cost of the protocol is defined as the number of bits the two players have to send to the Referee and the communication complexity of the task is defined as the minimum communication cost over all protocols that solve the task. In real world communication networks, very often the cost is rather calculated as the time one uses the communication channel, for example on the phone network. We note that these costs are interchangeable, provided that the communication channel has a specific maximum rate.
We define the time unit as the time to send a single bit over the communication channel, and then, in an optimal protocol, bits and communication time are equal, since one will always send one bit per time unit.
As custom, the time for local computations is ignored. Another resource one can study is the transmitted information, which instead of the number of bits sent, calculates the real bits of information about the inputs that the messages carry. For example, if Alice always sends the same, long message, independent of her input, then the communication time will be large, while the transmitted information will be zero, since no information about the input has been transmitted. Transmitted information is a resource that is important for privacy, when on top of having an efficient protocol, we want the Referee to solve the task without learning much about the players' inputs. One can define the transmitted information as the mutual information between the messages and the inputs and can upper bound it  with the logarithm of the number of different messages the players send. In any protocol, the transmitted information is at most the communication time, since one bit cannot carry more than one bit of information,
and hence, the bottleneck is always the time, as the largest quantity.

We can similarly define the resources for quantum protocols. The communication time is again the number of time units the protocol takes, where in a time unit at most one qubit can be sent in expectation. Here, we have added the ``in expectation'' since typically in quantum communications the qubits are realized by photons emitted by practical light sources and hence their mean number follows a Poisson distribution \cite{scarani2009security}. In the following, we also make this change to the classical model to make a more correct comparison, \emph{i.e.}, we allow one bit in expectation per time unit, which does not change the order of the communication time. We will also upper bound the transmitted information as the logarithm of the minimum dimension of the Hilbert space that contains all the possible quantum messages that are sent in the protocol. For example, if in a protocol Alice has as input an $n$-bit string $x$ and sends a message that contains $n/2$ qubits of the form $\ket{x_1}\otimes\ket{x_2}\otimes\ldots\otimes \ket{x_{n/2}}$ and another $n/2$ qubits in the state $\ket{0}\otimes \ldots \otimes \ket{0}$, then the communication time is $n$, while the transmitted information is $n/2$.

\emph{Euclidean distance of real vectors}. We now describe a fundamental communication task. Alice and Bob possess large data sets $x$ and $y$ respectively, which are unit vectors in $\mathcal{R}^n$. They would like to allow a Referee to check how similar their data is by estimating the Euclidean distance (for simplicity we define its square), $||x-y||_2^2 = \sum_{j=1}^n (x_j-y_j)^2$ (or equivalently the inner product, since $\langle x,y \rangle  = 1- ||x-y||_2^2/2$). We call this problem Euclidean Distance or ED. 

Alice and Bob can transmit their entire data to the Referee, but this is non-optimal. The idea is to send fingerprints of the data, which are much shorter but still allow the Referee to approximate their Euclidean distance within some additive constant. Classically, this problem requires Alice and Bob to send fingerprints of size $\Omega(\sqrt{n})$ \cite{ambainis1996communication,babai1997randomized,newman1991private,newman1996public}. We consider here that Alice and Bob do not have access to any shared randomness, otherwise the problem can be solved with only constant communication \cite{kremer1999randomized}. It is natural that parties do not \emph{a priori} have such shared resources, especially in large networks where the communication is between many different pairs of parties.

Quantum fingerprints can be exponentially shorter than the classical ones in this case. In particular, Alice and Bob create and send quantum fingerprints, namely
$\ket{\text{fin}_{x}} = \sum_{j=1}^{n}x_j\ket{j}$ and $\ket{\text{fin}_{y}}=\sum_{j=1}^{n}y_j\ket{j}$, respectively.
The Referee then estimates the Euclidean distance by performing a controlled swap operation on the fingerprints which outputs ``1'' with probability $1/2 + |\langle x,y \rangle |^2/2$ \cite{buhrman2001quantum}.
The Referee estimates this probability, and hence the Euclidean distance, within an additive constant $\epsilon$ with probability at least $1-\delta$, using a constant number of fingerprints equal to $\log (1/\delta)/\epsilon^2$. The communication time is $O(\log n)$, since Alice and Bob send a constant number of fingerprints and each fingerprint consists of $\log n$ qubits and can be sent in $\log n$ time units. The transmitted information is also $O(\log n)$, equal to the communication time. Hence, if we look at the ratio of the quantum over the classical resources, then both resources asymptotically go to zero as $n$ grows. Unfortunately, implementing these fingerprints with qubit systems is out of reach for current technologies for large $n$.

The notion of quantum fingerprints has been used in practice for the Equality problem \cite{arrazola2014quantum,xu2015experimental,guan2016observation}, where the inputs are binary strings and the Referee checks whether they are exactly the same or not. The Equality problem can be reduced with the help of error correcting codes to approximating the Euclidean distance between the two vectors within a constant factor and hence the previous protocol solves Equality with the same resources. Since most real data is represented as real-valued vectors, the Euclidean Distance problem is more pertinent than Equality, since it is rather improbable that two different sets of real-valued data will be exactly equal. Hence, here, we extend the use of the term quantum fingerprints to real-valued inputs, where we ask that fingerprints can be used to approximate the distance of the inputs and not check whether they are exactly equal or not.

\emph{Coherent state fingerprints for Euclidean Distance}. The coherent state mapping of \cite{arrazola2014coherent} led to a protocol for Equality with communication time $O(n)$ and transmitted information $O(\log n)$. This protocol therefore provides an exponential advantage in the transmitted information, at the expense of a quadratically worse performance in communication time compared to the classical protocol, for which the order of both resources is $\Omega(\sqrt{n})$.

\begin{figure}[h!]
\vspace{-0.1cm}
\includegraphics[scale=0.40]{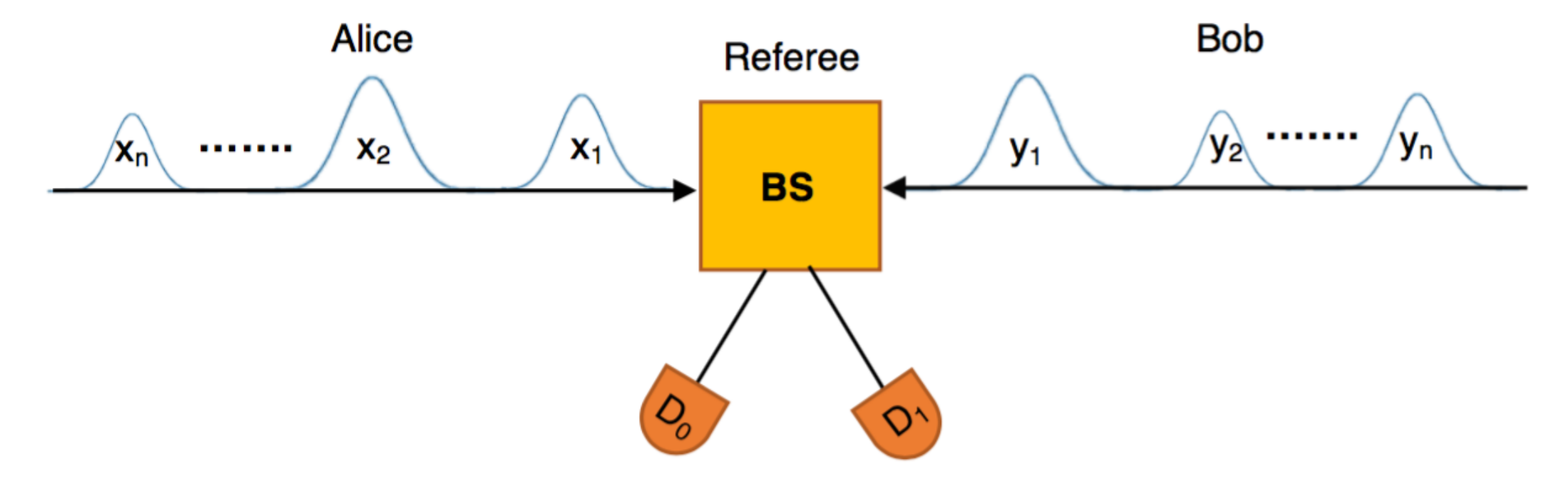}
\centering
\vspace{-0.5cm}
\caption{Alice and Bob send \textit{n} coherent pulses, with the $j^{\text{th}}$ pulse's amplitude determined by $x_j \alpha$ and $y_j \alpha$, respectively. The Referee interferes their states in a 50/50 BS and detects the output signals using single-photon detectors $D_0$ and $D_1$.}
\label{fig:cs}
\end{figure}

A schematic of the corresponding protocol for Euclidean Distance is shown in Fig.~\ref{fig:cs}. Alice and Bob's fingerprints are trains of $n$ coherent states sent to the Referee. Alice's state, $\ket{\alpha_{x}}$, is prepared by the displacement operator $\hat{D}_{x}(\alpha) = \exp(\alpha \hat{a}_{x}^{\dagger} - \alpha^* \hat{a}_{x})$ applied to the vacuum state, where $\hat{a}_{x} = \sum_{j=1}^{n}x_j\hat{b}_j$ is the annihilation operator of the fingerprint mode \cite{arrazola2014quantum}, and $\hat{b}_j$ is the photon annihilation operator of the $j^{\text{th}}$ time mode. Hence,
\begin{equation}
\ket{\alpha_{x}} = \hat{D}_{x}(\alpha)\ket{0} = \otimes_{j=1}^{n}\ket{x_j\alpha}_j,
\end{equation}
where $\ket{x_j\alpha}_j$ is a coherent state with amplitude $x_j \alpha$ occupying the $j^{\text{th}}$ mode. The mean photon number for the state $\ket{\alpha_{x}}$ is $\mu = \sum_{j} |x_j\alpha|^2 = |\alpha|^2$, independent of the input size.
Bob similarly creates the fingerprint $\ket{\alpha_{y}}$.


As shown in Fig.~\ref{fig:cs}, the Referee uses a $50/50$ beam splitter (BS) to interfere the incoming coherent states. This yields the output state for the $j^{\text{th}}$ time unit
\begin{equation}\label{output}
 \ket{\frac{(x_j + y_j)}{\sqrt{2}} \alpha}_{j,D_0}\otimes \ket{\frac{(x_j - y_j)}{\sqrt{2}} \alpha}_{j,D_1},
 \end{equation}
where the subscripts $D_0$ and $D_1$ denote the single-photon detectors placed at the output arms of the BS.

In previous works on Equality, the clicks of $D_1$ have been used for estimating how different the two fingerprints are. Then, since the expected number of clicks of the detector depends directly on its dark count probability, it is crucial to keep this probability very low. Here, we try to deal with this problem, by using the clicks from both detectors to construct a more robust estimator for the Euclidean distance that can also be used for Equality.

More precisely, let $Z^0_j$ and $Z^1_j$ be the binary random variables that are 1 with the probability with which $D_0$ and $D_1$ clicks respectively at the $j^{\text{th}}$ time unit, namely $p^0_j=1 - \exp(-\frac{\mu(x_j+y_j)^2}{2})\approx \mu\frac{(x_j+y_j)^2}{2}$, and $p^1_j=1 - \exp(-\frac{\mu(x_j-y_j)^2}{2})\approx \mu\frac{(x_j-y_j)^2}{2}$. Here, the approximation holds because we take $\mu$ to be typically small, and $x$ and $y$ are unit vectors in $\mathcal{R}^n$ and for large $n$ the terms $(x_j+y_j)^2$ and $(x_j-y_j)^2$ are typically in the order of $1/n$. The Euclidean distance ($\tilde{E}$) is equal to

\vspace{-0.5cm}
\begin{equation}\label{perfect}
\tilde{E} = 2 - \frac{1}{\mu}\mathbb{E}[\sum_{j=1}^n (Z^0_j-Z^1_j)].
\end{equation}
\vspace{-0.4cm}

The advantage of using statistics from both detectors comes from the fact that the Euclidean distance estimator depends now on the difference of the clicks of the detectors, and hence on expectation the number of dark counts cancels out, when we assume the dark count probabilities are the same for both detectors. We remark that this can be enforced by symmetrization procedures \cite{pappa2015nonlocality}, although in practice, since the symmetrization will not be perfect, the estimator will in fact depend on the square of the dark count probability, which is easier to keep low.

By Chernoff bounds, to estimate $\sum_{j=1}^n (Z^0_j-Z^1_j)$ within a constant factor $\epsilon$ with constant probability at least $1-\delta$, the number of fingerprints required is $O(\log (1/\delta)/\epsilon^2)$ \cite{upfal2005probability}. Hence, the overall communication time of the protocol is $O(n)$ while the transmitted information is $O(\mu\log n)$. Note that in each time unit, $\mu/n \ll 1$ photons are sent in expectation, thus satisfying our model's criterion of no more than one photon in each time unit.

In Table I we summarize in the first two rows the resources of the two protocols we have described for ED. The performance achieved with the coherent state fingerprint protocol is the same as the one achieved for Equality, \emph{i.e.}, exponentially better in transmitted information but quadratically worse in communication time. We describe now a quantum protocol that can perform better than a classical protocol in both resources.

\begin{table}[h!]
 \centering
\begin{tabular}{|c|c|c| }
 \hline
 & Comm. Time  & Trans. Info. \\ [1 ex]
 \hline
  Classical & $\Omega(\sqrt{n})$ & $\Omega(\sqrt{n})$   \\  [0.5 ex]
\hline
Coherent &  $O(n)$  & $O(\mu\log n)$ \\ [0.5 ex]
\hline
 Mux Classical & $\Omega(\frac{\sqrt{n}}{\log k})$ & $\Omega(\frac{\sqrt{n}}{\log k})$   \\  [0.5 ex]
\hline
 Mux Coherent & $O(\frac{n}{k})$ & $O(\mu\log n)$  \\  [0.5 ex]
 \hline
\end{tabular}
\caption{The order of the communication time and transmitted information for all classical and quantum protocols for Euclidean Distance described in this work.}
\vspace{-0.25cm}
\end{table}

\emph{Multiplexed coherent state fingerprints}. We extend both the classical and quantum communication models to allow Alice and Bob to have multiple communication channels with the Referee. In particular, Alice and Bob can use $k$ different channels, where in every communication time unit, they can send in expectation at most one bit or one photon in total over all $k$ channels.

First, the multiple channels reduce the classical communication by at most a $\log k$ factor, since we can simulate any multiple channel protocol with a single channel one with a $\log k$ overhead: for every bit sent through one of the $k$ channels, we send the same bit and the index of the channel in $\log k$ bits through the single channel.

In the quantum case, we take better advantage of the multiple channels and have an ED protocol with communication time of order $n/k$, while the transmitted information remains of order $\log n$.
The underlying reason is that most of the pulses sent are empty of photons and hence we can use the multiple channels to send in parallel many pulses, without sending more than one photon in expectation per time unit. More precisely, Alice and Bob divide their $n$ bit input into $k$ substrings, each of length $n/k$. They create coherent state fingerprints for each of the $k$ substrings and at each time unit they send $k$ pulses through the channels, one from each of the $k$ fingerprints. The Referee interferes the corresponding pulses as in the initial protocol, either by using $k$ sets of BS and detectors, or by time ordering the pulses and using a single set of BS and detectors. The communication time is now reduced by a factor of $k$. By choosing $k$ to be $\omega(\sqrt{n})$, we can make both resources of the quantum protocol asymptotically smaller than the best classical protocol. The expected number of photons in each time unit is $\mu k/n$, which for large enough $n$ and since $k$ is asymptotically smaller than $n$ can be made $<1$, hence satisfying the no more than one photon per time unit constraint.

One way to implement the above protocol is by using $k$ physical channels. This is the case for backbone communication networks, where nodes are connected via a large number of channels. Another way could be to employ all-optical orthogonal frequency division multiplexing (OFDM), an advanced classical multiplexing technique that has recently been adapted for performing high-rate quantum key distribution \cite{bahrani2015orthogonal} (details in Appendix).

Figure~\ref{fig:mux1} illustrates an abstract implementation based on multiplexing. Alice and Bob create coherent state fingerprints for each of the $k$ substrings and at each time unit they multiplex the corresponding set of $k$ pulses and send the output signal to the Referee. The Referee demultiplexes the signals from Alice and Bob and then interferes each pair of pulses through a BS. The protocol proceeds similarly for all $n/k$ time units and the Referee estimates the Euclidean distance as described before.

\begin{figure}[h!]
\includegraphics[scale=0.30]{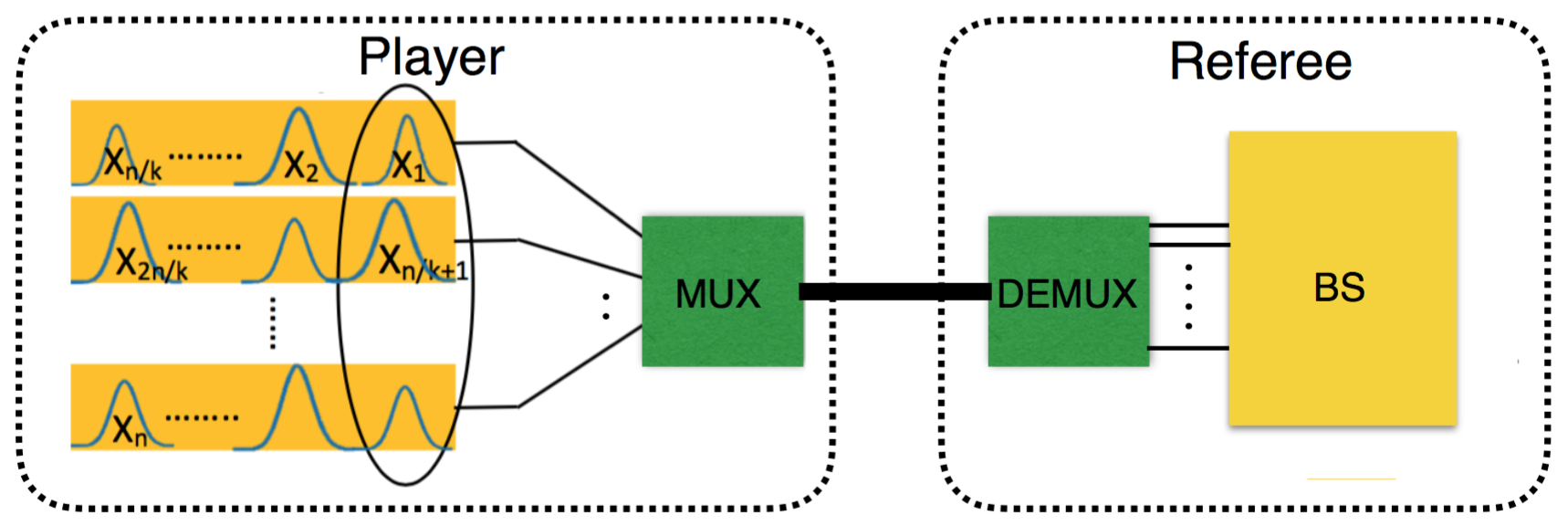}
\centering
\caption{ED protocol with multiplexed coherent state fingerprints. Alice's pulses are multiplexed by MUX and sent to the Referee who demultiplexes them with DEMUX and interferes them with the pulses received from Bob using the BS.}
\vspace{-0.4cm}
\label{fig:mux1}
\end{figure}

The last two rows of Table I compare the classical and quantum multiplexed protocols for ED. This is the first example of a communication model and a task, for which an in principle realistic quantum protocol is asymptotically more efficient both in the communication time and the transmitted information than any classical protocol.

\begin{figure}[h!]
\includegraphics[scale=0.30]{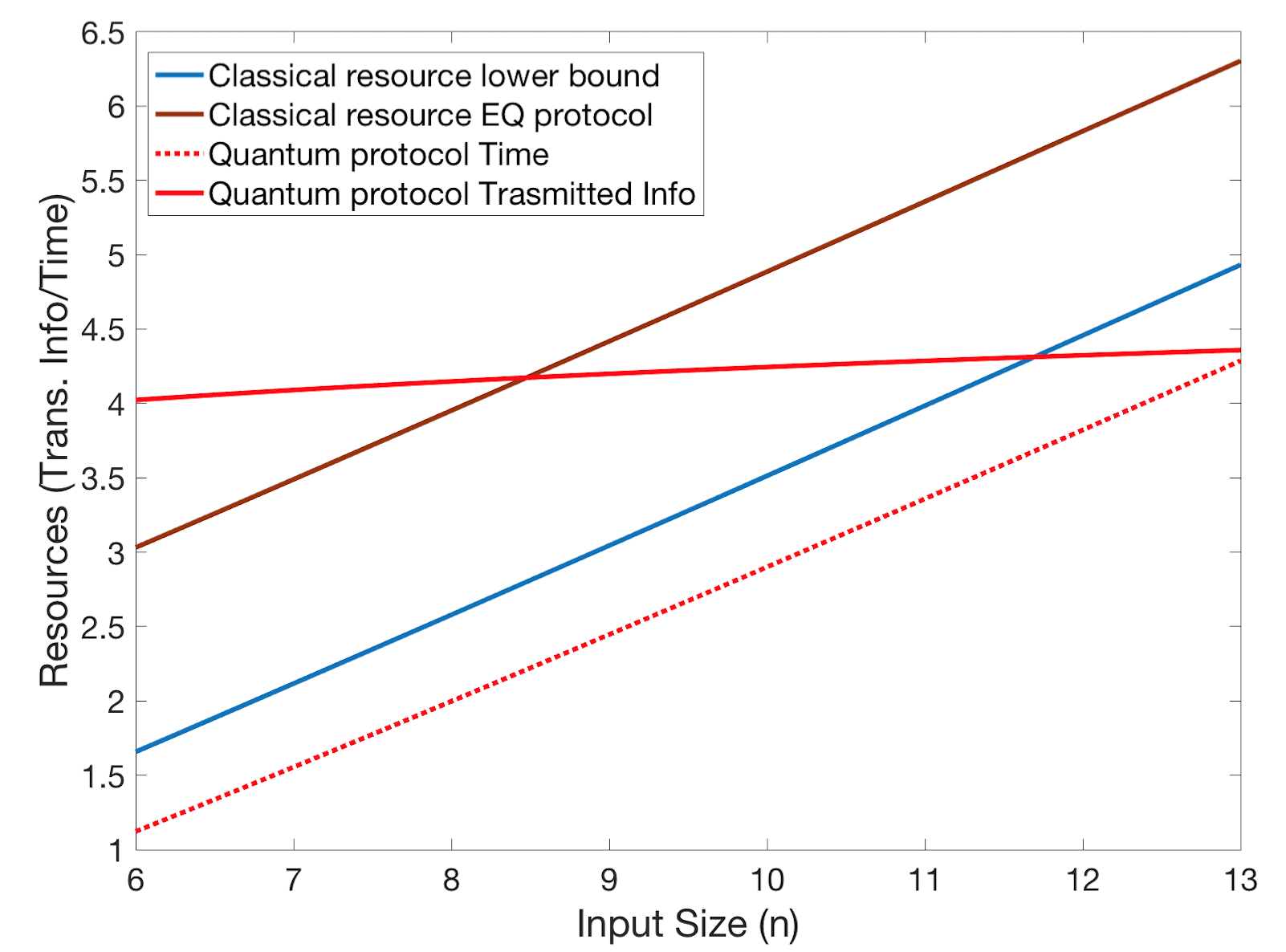}\centering
\vspace{-0.3cm}
\caption{Log-log plot for transmitted information and communication time vs input size $n$, comparing the classical lower bound, the best known classical protocol and our multiplexed protocol, for $k=20\sqrt{n}\log n$, $\mu\sim 100$ and $\nu = 0.99$. Our protocol approximates ED within $0.2$ with error  $\delta \leq 10^{-6}$.}
\label{fig:comb}
\vspace{-0.4cm}
\end{figure}

\emph{Performance analysis}. We consider some standard experimental imperfections, including BS interferometer visibility $\nu$ and detector dark count probability $p_{d}$.
Errors may be due to multiplexing and demultiplexing as well but as these depend on the specific implementation we do not consider them here. Moreover, we assume that all of Referee's detectors have the same parameters, in particular that they have equal dark count probabilities.

With the above imperfections we recalculate the probability of a click in $D_0$ and $D_1$ and show in Appendix that
\vspace{-0.2cm}
\begin{equation}
\tilde{E} = 2 - \frac{1}{\mu(2\nu-1)}\mathbb{E}[\sum_{j=1}^n (Z^0_j-Z^1_j)].
\vspace{-0.2cm}
\end{equation}

The main source of error comes from the error in estimating the expectation of the detectors clicking. This error is handled by repeating the protocol a constant number of times and using Chernoff bounds. Another source of error comes from estimating the experimental parameters, i.e. the BS visibility and the mean photon number. Note that the expectation of the difference in counts of the two detectors does not depend on the dark count probability, provided it is the same for 0 and 1.

In Fig.~\ref{fig:comb} we show the transmitted information and communication time as a function of $n$, for our multiplexed protocol with $k$ of the order of $\sqrt{n}$, and compare its performance with the classical lower bound as well as with the best known classical protocol for Equality. The analytical expressions for all protocols are provided in Appendix. We see first, that if we are only interested in the communication time, which is often the case, then our protocol outperforms the classical limit even for small $n$ and by consequence for small number of multiplexed channels which can be feasible in practice. Moreover, for large enough $n$, our protocol outperforms the classical limit for both resources. For current parameters, the number of channels needed is in the order of $10^5$, which may not be realistic. By improving the experimental parameters it may be possible to decrease this number.

\emph{Discussion}. A noteworthy feature of our protocol is that Alice and Bob do not need a memory to store their inputs and they do not perform global operations on them. In other words, our protocol works also in the {\em streaming} scenario, where Alice and Bob receive their inputs one bit at a time \cite{ams}. We note that this is not the case neither for the Equality protocol, where an error correcting code needs to be applied to the entire input string, nor for the qubit protocol where the fingerprint is encoded in a superposition of $\log n$ qubits. It will be interesting to further explore this scenario for efficient quantum communications. More generally, expanding the family of distributed tasks in the coherent state communication model studied in this work is important for demonstrating in practice quantum superiority in a network setting.\\


We thank Juan-Miguel Arrazola for useful discussions. We acknowledge financial support by the ERC project QCC, the ANR projects COMB and QRYPTOS, the Region Ile-de-France DIM Nano-K project QUIN, and the Partner University Fund project CRYSP.\\

\bibliography{fingerprints}

\begin{widetext}
{\center \bf Appendix}

\emph{Description of OFDM approach for multiplexed coherent state Euclidean distance}. We provide some details for employing all-optical orthogonal frequency division multiplexing (OFDM) in our multiplexed coherent state fingerprint framework. In this case, Alice and Bob create coherent state fingerprints for each of their $k$ substrings in orthogonal frequency subcarrier modes that are generated using frequency offset locked laser diodes (alternatively, a pulsed laser source such as a mode-locked laser can be used for this purpose, as shown in \cite{bahrani2015orthogonal}). At every time unit, the corresponding pulse from each of the $k$ fingerprints is multiplexed in a $k\times 1$ OFDM encoder and the output signal is sent to the Referee.

The frequency separation between any two adjacent subcarriers in the OFDM scheme is $\omega_{j+1} - \omega_{j} = \Delta f$ and the encoded OFDM signal has a pulse width of T = $1/\Delta f$.
At the first time unit $t_1$, Alice's coherent pulses for each of the $k$ substrings are $\{\ket{x_1\alpha}_1,\ket{x_{\frac{n}{k}+1}\alpha}_1,..,\ket{x_{\frac{(k-1)n}{k}+1}\alpha}_1 \}$. They get encoded in the OFDM signal $\hat{E}_{1}(t)$ (subscript denotes time):
\begin{equation*}
\hat{E}_{1}(t) = \sum_{j=1}^{k}e^{-\frac{|x_{ \frac{(j-1)n}{k}+1}|^2\mu}{2}} e^{x_{\frac{(j-1)n}{k}+1}\alpha\hat{a}_{j}^{\dagger}} e^{i \omega_{j} t} = \sum_{j=1}^{k}\hat{A}^j_{1} e^{i \omega_{j} t}
\end{equation*}
for $0<t<$ T, with the $j^{\text{th}}$ subcarrier frequency given by $\omega_{j}$ = $\omega_{0}$ + 2$\pi j \Delta f$. Bob employs the same multiplexing technique to prepare his OFDM signal.

Once the OFDM signals from Alice and Bob for the time step $t_1$ reaches the Referee, he decodes them via an Optical Discrete Fourier Transform (ODFT) on the input signal $\hat{E}_{1}(t)$.
The output circuit to decode the $q^{\text{th}}$ subcarrier signal (for $q = {1,..,k}$) at time step $t_1$ is:
\begin{equation*}
\hat{D}_{1}^{q}(t) = \frac{1}{k}\sum_{j=1}^{k}\hat{E}_{1}(t - (j-1)T_c)e^{i 2 \pi (j-1) (q-1)/k} = \hat{A}^q_{1} e^{i \omega_{q} t}
\end{equation*}
where $T_c = T/k$ and we used the orthogonality condition $\Delta f$ = $1/T$. A typical duration for the OFDM signal is T = 100 ps \cite{bahrani2015orthogonal}.

The advantage of this technique is that because of the orthogonality of the employed subcarriers, these do not interfere with each other despite overlapping sidebands between adjacent carriers, leading to an efficient demultiplexing; however, the number of supported subcarriers in practice currently remains quite low.\\

\emph{Euclidean distance expression with experimental imperfections}. The Euclidean distance between the data sets $x$ and $y$ is $\tilde{E} = \sum_j(x_j - y_j)^2 = ||x - y||^2$. We prove directly the case with interferometer visibility $\nu$.
Let $Z^0_j, Z^1_j$ be the binary random variables that take the value 1 with probability $\Pr[\mbox{click in } D_0]$ and $\Pr[\mbox{click in } D_1]$ respectively, for the $j^{\text{th}}$ time unit. We have assumed that all detectors have the same parameters so even in the case the Referee uses a different set of detectors for each time unit $j$, the probabilities are the same.
These probabilities are $ p_j^{D_0} \approx \nu\frac{(x_j+y_j)^2}{2}\mu + (1-\nu)\frac{(x_j-y_j)^2}{2}\mu + p_{d}$; and $ p_j^{D_1} \approx \nu\frac{(x_j-y_j)^2}{2}\mu + (1-\nu)\frac{(x_j+y_j)^2}{2}\mu + p_{d}$, where $p_d$ is the dark count probability for both detectors. The expectation value of their difference over $n$ is,
{\small
\begin{align} \label{eq:6}
\mathbb{E}[\sum_{j=1}^n (Z^0_j - Z^1_j)] = \frac{2\nu-1}{2}\mu \big(||x + y||^2-||x - y||^2\big)
\end{align}
}

From this expression and using the fact that $||x+y||^2-||x-y||^2=4(1-||x-y||^2/2)$, we obtain
\begin{equation} \small \label{eq:8}
\tilde{E} = ||x-y||^2 = 2(1-\frac{1}{4}(||x+y||^2-||x-y||^2)) =
2 - \frac{1}{\mu(2\nu-1)}\mathbb{E}[\sum_{j=1}^n (Z^0_j - Z^1_j)].
\end{equation}


Equation~(\ref{eq:8}) shows that the error in the estimation of the Euclidean distance comes from two different sources: first, the estimation of the mean value of the sum $\sum_j (Z^0_j - Z^1_j)$. For this, using the Chernoff bound, we can deduce that if the Referee wants to estimate this within a small constant factor $\epsilon$ with probability at least $1-\delta$, the number of samples required is $\leq \frac{3}{\epsilon^22\mu(2\nu-1)}\log (\frac{1}{\delta})$~\cite{upfal2005probability}; second, the error in the parameter estimation of $\mu$ and $\nu$, which in general depends on the experimental setup but can be considered very small.\\



\emph{Details of performance analysis}. The plot in Fig.~\ref{fig:comb} compares the transmitted information $(I)$ and communication time $(T)$ vs. the data input size $n$ for  the classical lower bound; the best classical protocol; and the quantum multiplexed coherent state protocol.  We fix that the protocol estimates ED within a constant factor $\epsilon = 0.2$ with error probability $\delta \leq 10^{-6}$.

\underline{Classical lower bound}:
We use the lower bounds for the Equality problem \citep{ambainis1996communication,babai1997randomized,newman1996public} to provide a lower bound for ED.

Suppose we have a classical ED protocol for input size $n$ that approximates the distance within a fixed $\epsilon$ with probability at least $1-\delta$. To construct a protocol for Equality, we choose the error-correcting code (ECC) that amplifies the $n$-bit inputs $x$ and $y$ to $m$-bit codewords $E(x)$ and $E(y)$ respectively, with the minimum distance across being $d>2\epsilon$. Then, we use the ED protocol on the codewords $E(x)$ and $E(y)$, and have:
\begin{equation} \label{eq:10}
\tilde{E} \begin{cases}
\leq \epsilon & \text{ if } x= y\\
\geq d - \epsilon > \epsilon & \text{ if } x\neq y
\end{cases}
\end{equation}
Hence, this guarantees solving the Equality on $x$ and $y$ with probability $\geq 1-\delta$.

Hence we get a lower bound for ED as $T_{cl} = I_{cl} = \big((1-2\sqrt{\delta})\sqrt{\frac{n}{2\log2}} - 1\big)/\log k$ \cite{babai1997randomized,guan2016observation}, which is shown in Fig.~\ref{fig:comb}.

\underline{Classical protocol for Equality}: Here we also plot the best classical protocol that solves Equality, which uses $2\sqrt{n} + 1$ bits and succeeds with probability $1-\delta = 3/4$. To get the desired $\delta\leq 10^{-6}$, the protocol is repeated 10 times. Thus $T_{cp} = I_{cp} = \big(20\sqrt{n} + 10\big)/\log k$ \cite{babai1997randomized}.

\underline{Quantum multiplexed coherent state protocol}: Our protocol bounds the ED within $\epsilon$ with probability $1-\delta$ and has transmitted information
$I_{qp}= \frac{3}{\epsilon^22\mu(2\nu-1)}\log(\frac{1}{\delta})\mu\log_2n$; and communication time $T_{qp} = \frac{3}{\epsilon^22\mu(2\nu-1)}\log (\frac{1}{\delta})n/k.$ 

In order to provide an advantage both in the transmitted information and the communication time, we take $k = \frac{3}{\epsilon^22\mu(2\nu-1)(1-2\sqrt{\delta})}\log (\frac{1}{\delta})\log_2(\sqrt{n})\sqrt{2\log 2n}$. Hence, the number of channels $k$ scales as $O(\sqrt{n}\log n)$.
For the plot, we consider $k=20 \sqrt{n}\log n$ and the experimental parameters to be $\mu\sim 100$ and $\nu = 0.99 \pm 0.005$ \cite{xu2015experimental}. 

\end{widetext}

\end{document}